\documentclass[journal]{IEEEtran}
\usepackage{amsmath}
\usepackage{tikz}
\usepackage{enumerate}
\usepackage{pgfplots}
\usepackage{subfigure}
\usepackage{amsthm}
\usepackage{siunitx}
\usepackage{multirow}
\usepackage{flushend}
\usepackage{caption}
\usepackage{subcaption}
\usepackage{graphicx}

\usepackage{booktabs}

\usepackage[ruled,vlined]{algorithm2e}

\usepackage{tablefootnote}

\usepackage{hyperref}

\ifCLASSINFOpdf
\else
\fi
\begin{document}
%
\title{Round Trip Time (RTT) Delay in the Internet: Analysis and Trends}
%
%
%

\author{Gonzalo~Mart\'inez,~\IEEEmembership{Student Member,~IEEE,}
        Jos\'e~Alberto~Hern\'andez,~\IEEEmembership{Member,~IEEE,}
        Pedro~Reviriego,~\IEEEmembership{Senior~Member,~IEEE,}
        ~and~Paul~Reinheimer
\thanks{G. Mart\'inez and J.~A.~Hern\'andez are with the Department
of Telematic Engineering, Universidad Carlos III de Madrid,
28911, Legan\'{e}s, Spain.}
\thanks{P. Reviriego is with the Department of Telematic Systems Engineering, Universidad Polit\'{e}cnica de Madrid, 28040, Madrid, Spain.}
\thanks{P. Reinheimer is with WonderProxy, Waterloo, Ontario, Canada.}
}

%
%

\markboth{Article accepted in IEEE Network Magazine}%
{Martinez \MakeLowercase{\textit{et al.}}: RTT Delay in the Internet}
%



\maketitle

\begin{abstract}

Both capacity and latency are crucial performance metrics for the optimal operation of most networking services and applications, from online gaming to futuristic holographic-type communications. Networks worldwide have witnessed important breakthroughs in terms of capacity, including fibre introduction everywhere, new radio technologies and faster core networks. However, the impact of these capacity upgrades on end-to-end delay is not straightforward as traffic has also grown exponentially. This article overviews the current status of end-to-end latency on different regions and continents worldwide and how far these are from the theoretical minimum baseline, given by the speed of light propagation over an optical fibre. We observe that the trend in the last decade goes toward latency reduction (in spite of the ever-increasing annual traffic growth), but still there are important differences between countries. 
\end{abstract}


%
\IEEEpeerreviewmaketitle

\section{Introduction}

\IEEEPARstart{O}{ver} the years, backbone networks have been increasing their capacity steadily, thanks to the continuous growth in the capacity-reach product of optical networks. Indeed, Nielsen’s law of Internet bandwidth, which states that  the speed of a high-end user’s Internet connection should grow by 50 percent each year, has been very accurate in the past decades. This has been possible mainly thanks to the advances in optical fiber transmission, which has increased from a single channel at 10~Gb/s per fiber in the late 1990s to several tens of channels at 100~Gb/s per fiber in 2010s~\cite{xia}.



As a matter of fact, the NSFnet was brought online in 1986 taking over the original ARPANET. It offered backbone capacity values of 56~Kb/s, a number that was later upgraded to 1.5~Mb/s in 1988, 45~Mb/s in 1991, and 145~Mb/s in 1994, right before its privatization~\cite{hist_nsf}. The Abilene project was then established in 1999, inteconnecting 230 nodes, mostly universities. Originally, the Abilene network backbone had a capacity of 2.5 Gb/s, and was upgraded to 10 Gb/s by February 2004. Abiline was officially retired in 2007, becoming Internet2, and boosting its capacity to 100~Gb/s.

The bandwidth increase evolution can also be analysed from the point of view of Ethernet standards. Fast Ethernet standard was approved in 1995 (IEEE 802.3u), Gigabit Ethernet in 1999 (IEEE 802.3z), 10~GbE in 2002 (IEEE 802.3ae), 100~GbE in 2010 (IEEE 802.3ba), 200 and 400~GbE in 2017 (IEEE 802.3bs) and at present, 800~GbE and 1.6~TbE are on their way to become a standard in the forthcoming years (IEEE P802.3df Ethernet Task Force). 

The access network (last mile) has also witnessed impressive boosts in speed, going from few Kb/s of dial-up over twisted pair in the 1990s, to few Mb/s in the 2000s via HFC and DSL technologies, and reaching multiple gigabit per second connectivity as of today. On the ITU side, G-PON, XG-PON, XG(S)-PON and 50G-PON standards (ITU G.984, G.987, G.9807 and G.9804.3) allow up to 50~Gb/s shared between 16 or 32 users typically, while NG-PON2 (ITU G.989) provides a 4x and 8x bandwidth capacity increase with respect to XG-PON. On the IEEE side, EPON (IEEE 802.3ah), 10G-EPON (IEEE 802.3av), 25/50G-EPON (IEEE 802.3ca) and the forthcoming Super-PONs (IEEE P802.3cs) also allow tens of Gb/s per PON tree (typically less than 64 households).

However, public and private investment on backbone networks have been uneven among countries, with developed countries going much faster in the deployment of fiber based networks, even reaching the final user with Fiber to the Home deployments. In general, high-capacity networks also enable lower packet transmission times since they reduce several sources of delay. For example, in links above 10~Gb/s, the transmission of a 1500~byte packet takes less than $1.2~\mu s$, while traversing a 1~Km link over fiber takes $5~\mu s$. Thus, ultimately, latency, in the absence of queuing should be approaching the lowest bound determined by the speed of light, i.e. $3.33~\mu s/Km$ for air propagation and $5~\mu s/Km$ for silica fiber or cable transmission media. 

Indeed, reducing end-to-end latency has a critical impact in the forth-coming services and applications. Human senses demand latency values in the ranges of 1 to 100~ms \cite{TolerableDelay} and 1~ms latency is the requirement for the tactile Internet \cite{TactileInternetDelay}. Other non-human applications, in particular those under the group of Ultra-Reliable Low-Latency Applications (URLLC) defined for 5G, clearly impose tight latency requirements to properly operate. Examples include tele-surgery, remote robot operations for the connected industry 4.0, connected cars or data centers to name a few~\cite{modyano_ieeeproc}.

In this article, we evaluate the current status of end-to-end latency across different regions of the globe, using real network ping statistics provided by the WonderNetwork measurement platform\footnote{See WonderNetwork's ping statistics, at https://wondernetwork.com/pings, last access April 2023.}. The dataset contains over 9 million measurements between over 100 monitoring sites, between 2014 and 2022. The main goals of this paper are then (a) to analyse the current status of end-to-end latency worldwide between monitoring stations located in major cities, (b) to study how end-to-end latency has evolved in the last decade, and (c) to understand the differences between regions worldwide regarding Internet connectivity latency. 

The remainder of this work is organised as follows: Section~\ref{sec:delay} describes the main sources of delay in packet switching networks. Section~\ref{sec:methodology} reviews the data sources and sets used in the article. Section~\ref{sec:results} presents the analysis and the results obtained from the RTT datasets worldwide, while section~\ref{sec:applicability} shows an example use case of application of the regression models. Finally, section~\ref{sec:discussion} concludes this work with its main findings and conclusions.

\section{Understanding the Sources of Delay}
\label{sec:delay}

In the Internet, a data packet has to traverse multiple links and nodes (cables, fibres, routers, switches, etc) to reach its destination and each step along the way adds some delay~\cite{KuroseRoss21}. This leads to different delays across regions and continents and wide range measurement datasets are needed to understand Internet delay~\cite{RTTInternet},\cite{RTTInternet2}.

\begin{table*}[!htbp]
\centering
\caption{Sources of delay and trends}
\label{tab:delaytypes}
\begin{tabular}{@{}lll@{}}
\toprule
Type   & Typical values  & Trend                                                          \\ \midrule
Transmission & Less than $1 \mu s$ for packets on links above 10 Gb/s & Reducing since link capacities keep growing \\
Queueing   & Typically 1 to 4 packet transmission times ahead in queue at loads above 75\% & Reduction again since link capacities keep growing  \\
Processing & Less than $1\mu s$ for high-performance switches and routers & Reduction since hardware acceleration keeps growing \\
Propagation  & $5\mu s$ per Kilometer  & Fixed, depends on distance and speed of light over silica fiber \\ \bottomrule
\end{tabular}
\end{table*}

In general, the main sources of delay are (see also Table~\ref{tab:delaytypes}):
\begin{enumerate}
    \item Processing delay: this is the time taken by a router or switch to decide the correct output interface for that packet. The tasks involved include checking layer-2 CRC, inspecting destination address, checking the internal lookup table, decreasing TTL of the IP header, etc.
    \item Transmission delay: this accounts for inserting each individual bit of the packet in the transmission media. This value depends on the bitrate of the interface card and packet length.  
    \item Queuing delay: there may be other packets that are waiting to be sent on the outgoing port such that our packet has to be buffered and wait for service.
    \item Propagation delay: Once the packet has been transmitted on a link, it has to reach the other end of the link. This time depends on the distance between the link endpoints and the speed of light on that transmission media.
\end{enumerate}

Processing delay used to be a significant contributor to overall delay with values of hundreds to thousands of $\mu s$ for software implementations depending on the complexity of the processing done on the packets \cite{ProcessingDelay}. However, it has been consistently reduced due to the increase in speed and performance of processors and to the use of hardware acceleration mechanisms for the most complex functions or even full hardware implementations for high speed routers\footnote{Details on the latency of several switches that achieve sub-$\mu s$ delay are available at \url{https://network.nvidia.com/pdf/whitepapers/FTTC_10GbE_Report_FINAL.PDF}, last access April 2023.}. Today the processing delay of a high speed router can be in the order of hundreds of nanoseconds for hardware implementations up to a few microseconds on software implementations and is expected to keep reducing over time in the future. 

Similarly, transmission delay is also a diminishing factor in network delay as the size of packets has roughly stayed the same in the last decades while the link speeds have increased by several orders on magnitude. This has led to transmission delays of $1.2~\mu s$ for a maximum size Ethernet packet on a 10 Gb/s link and of only $30~ns$ for state of the art 400 Gb/s links. 

The queuing delay, is heavily dependent on the network load, that in many cases is low as networks tend to have extra capacity to be able to keep up with traffic growth over time. In more detail, the M/M/1 model estimates that the average waiting time in queue for a packet that arrives at a network interface card (NIC) operating at 50\% load is of one packet ahead (i.e. another $1.2~\mu s$ at 10~Gb/s or $30~ns$ at 400~Gb/s). In the more generic GI/GI/1 model, this number may be multiplied by 10x or 100x at most under the assumption of high burstiness in the packet arrival process (using the Kingman's formula~\cite{kingman_1961}), which would still be only a few microseconds per IP hop. Therefore in many scenarios queuing delay is also limited except maybe during short periods of time where traffic peaks occur.

Finally, the propagation delay depends on the distance and transmission media. On free space, the signals propagate at the speed of light so introducing a delay of approximately  $3.33~\mu s$ per kilometer. Instead most guided media such as copper or silica fiber have lower propagation speed and introduce a delay that is around  $5~\mu s$ per kilometer. 

In conclusion, the speed of light is the ultimate lower bound in delay since the other sources of delay can be reduced with technological advances (see Table~\ref{tab:delaytypes}). However, the distance between today's latency and that minimum theoretical value varies for different parts of the globe. The next section quantitatively analyses this gap and its main reasons.

\section{Methodology}
\label{sec:methodology}

\begin{figure*}[!htbp]
\centering
\includegraphics[width=0.9\textwidth]{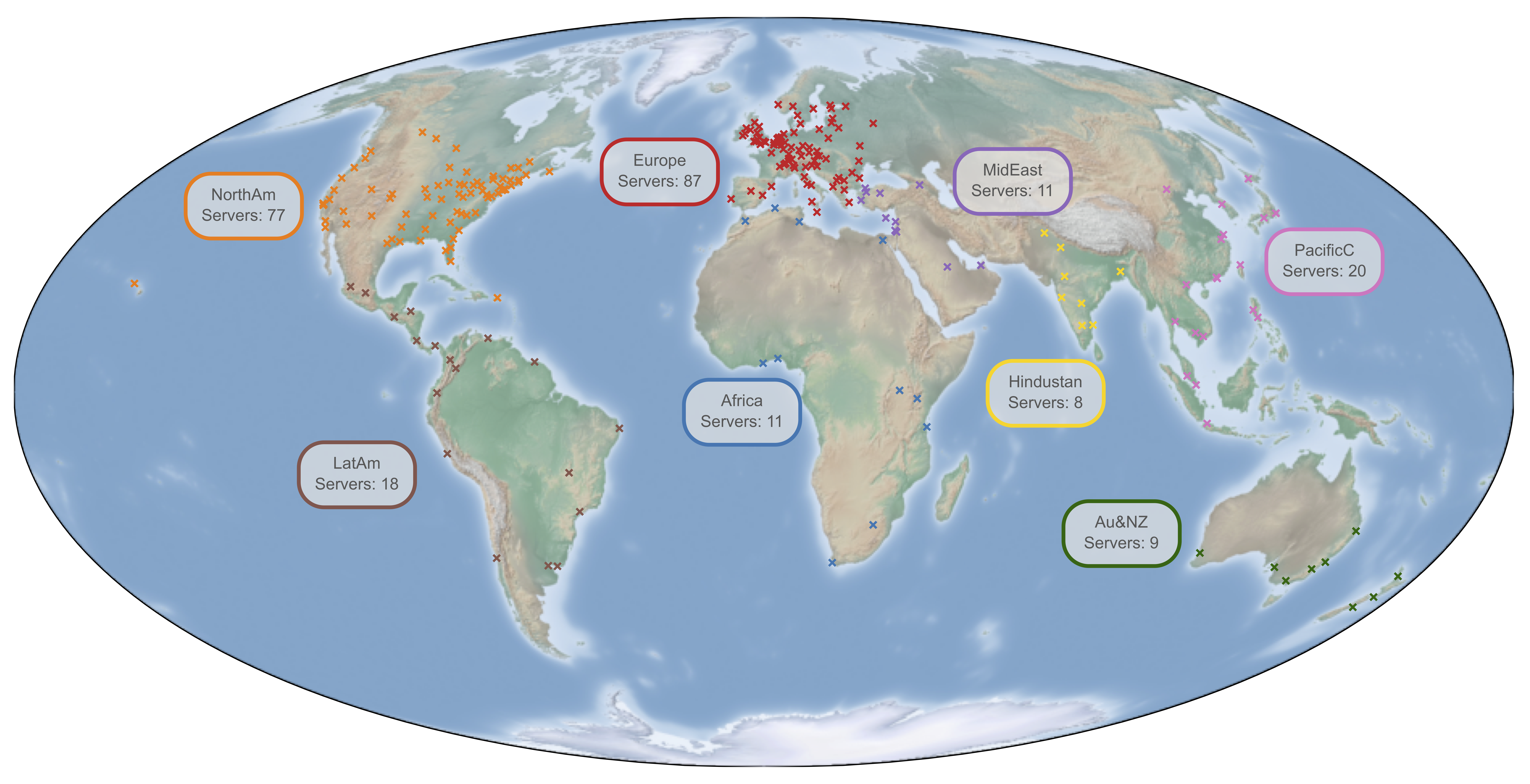}
\caption{Location of monitoring stations around the world.}
\label{fig_DS1}
\end{figure*}

The following experiments shall use a dataset that comprises ping measurements (two-way delay, RTT) between pairs of monitoring stations widely distributed across the globe. Fig.~\ref{fig_DS1} shows the locations of such monitoring stations, clustered into eight main regions. This dataset has been kindly provided by WonderNetwork, and part of it can be accessed directly from their website\footnote{WonderNetwork Ping statistics, url: https://wondernetwork.com/pings, last access April 2023.}. 

The measurements include the minimum, average, maximum and standard deviation between every pair of monitoring stations, at different times of the day, and have been collected between 2014 and 2022. We have arranged the RTT measurements into into eight groups or sub-datasets, namely, NorthAm (North America, 77 monitoring stations), Europe (both Eastern and Western Europe, 87 stations), Au\& NZ (Australia and New Zealand, 9 stations), LatAm (Latin America, 18 stations), Africa (11 stations), MidEast (Middle East, 11 stations), Hindustan (India, Pakistan, Bangladesh, 8 stations), and PacificC (Japan, Singapore, Malaysia, Indonesia, China\textcolor{red}{,} Thailand, Vietnam, Philippines, South Korea and Cambodia, 20 stations). In addition, for each source-destination RTT measurement pair, we have computed the shortest geographical distance (in Kilometers) obtained from GPS coordinates (latitude and longitude). It is worth noticing that the number of monitoring stations and servers is larger (therefore richer RTT datasets) in North America and Europe (77 and 87 servers respectively) than in other subsets of data.

\section{End-to-end delay Analysis}
\label{sec:results}

\subsection{Delay versus location: not all continents are equal}

\begin{figure}[!htbp]
\centering
\includegraphics[width=0.9\columnwidth]{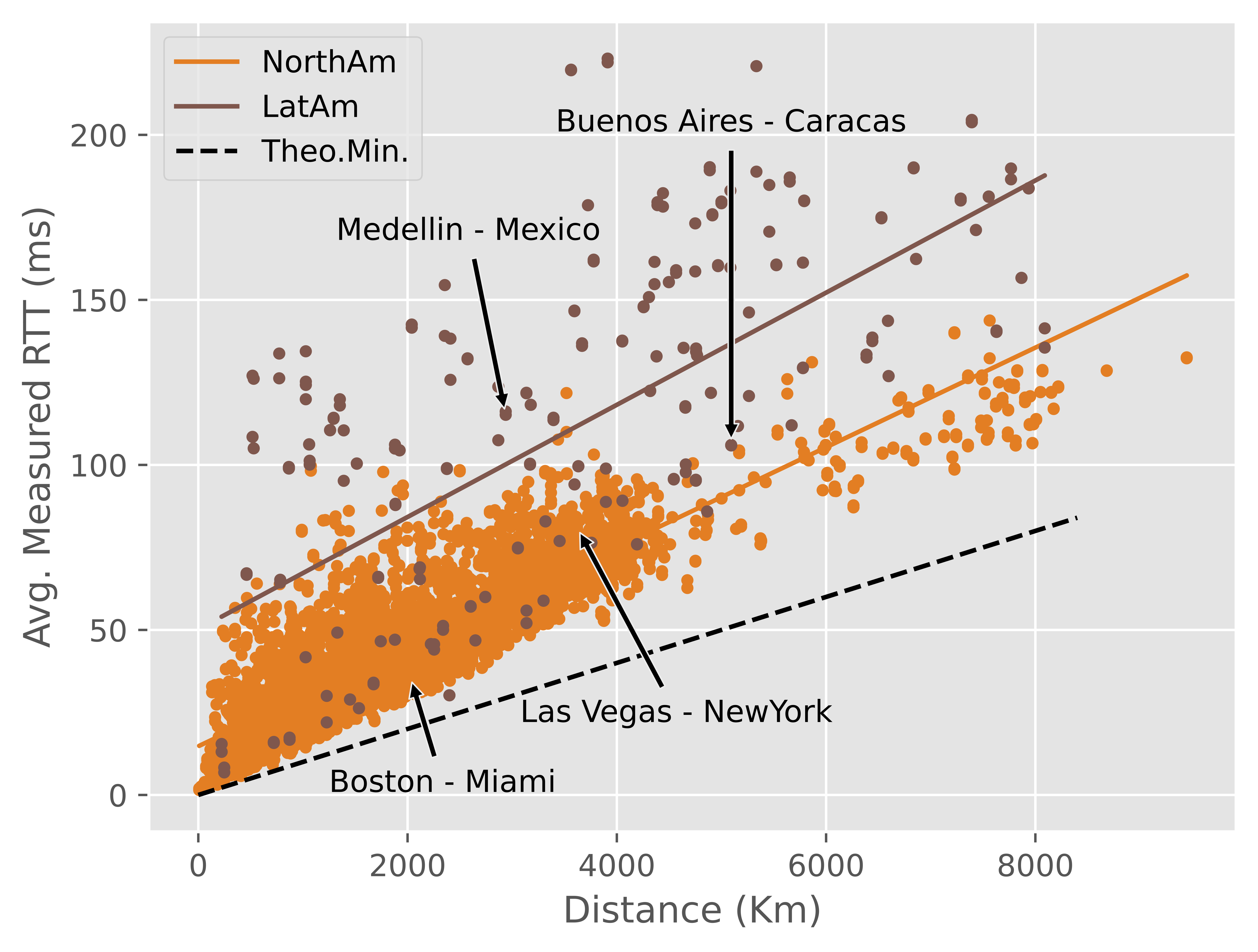}
\caption{Dataset measurements along with regression lines for NorthAm and LatAm}
\label{fig:datapoints}
\end{figure}

Fig.~\ref{fig:datapoints} shows a scatter plot of different datapoints for the NorthAm and LatAm datasets, year 2022. The coordinates of each dot correspond to the distance in Km (x-axis) and the observed RTT in millisecs (y-axis) between pairs of monitoring stations (orange for NorthAm and brown for LatAm); each dot represents the average RTT of a pair of cities, for example Boston-Miami, Las Vegas - New York, Buenos Aires - Caracas, etc. Along with the measurements, the figure shows regression lines of the average RTT values as a function of distance, for the two datasets. Finally, the dashed line represents the theoretical minimum baseline, which considers the propagation delay only between any two points using the shortest geographical distance. Such a baseline has null intercept and $0.01~ms/Km$ slope as it follows from propagation delay across a fiber (that is $5~\mu s/Km$ one-way, or $0.01~ms/Km$ two-way propagation delay). This represents the best case of a direct point-to-point fiber interconnecting two points over the shortest geographical path. Obviously, this theoretical straight fiber connection is often not possible due to geographical constraints like mountains, sea, desert, etc. 

\begin{figure*}[!t]
\centering
\includegraphics[width=0.75\textwidth]{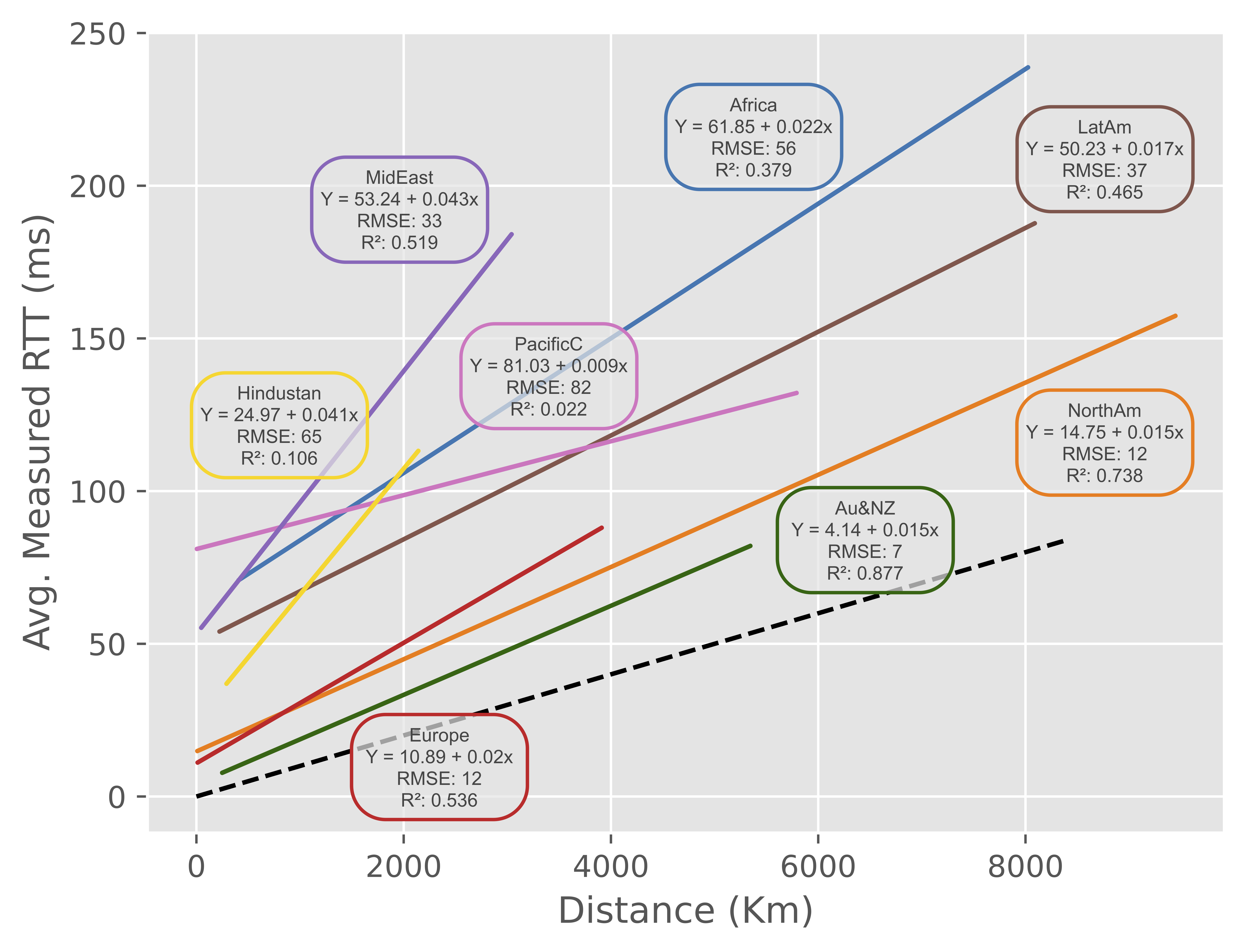}
\caption{Linear regression models for continents: min. RTT vs distance (year 2022).}
\label{fig_digitalgap_continents}
\end{figure*}

Fig.~\ref{fig_digitalgap_continents} further shows only  regression lines for measured average RTTs between different monitoring stations located in major cities world-wide (not the dot-datapoints). The linear regression models have been obtained from data measurements using classical linear regression techniques available in any Machine Learning (ML) library. ML libraries also provide exhaustive model information, including the Root Mean Square Error (RMSE), the coefficient of determination ($R^2$) and the actual regression line model, also depicted in the figure. 

As shown in Fig.~\ref{fig_digitalgap_continents}, North America, Au\&NZ and Europe show the closest latency values to the theoretical minimum baseline. For instance, ping values between monitoring stations located in North-America cities have a slope of $0.015~ms/Km$ (that is 1.5x above the theoretical minimum) and only 14~ms intercept. European cities experience a slope of $0.02~ms/Km$ (i.e. twice the theoretical minimum) and only 10~ms intercept, showing less delay that NorthAm cities for short distances (below 500 Km). Also Au\&Nz measurements, although with only 9 monitoring stations, show close-to-optimal latency values. Finally, other continents show far-from-optimal latency values, given the difficulty to deploy straight fiber lines across deserts, mountains and sea, needing longer distances between cities. For example, the lack of links traversing the Sahara desert forces connections between Tunis and Lagos to circumvent the continent leading to an RTT delay of about 150~ms, which is three times larger than the theoretical minimum of 50~ms for these two cities. This experiment reveals the current digital gap (in terms of latency) between continents as of year 2022. 

In addition to the regressed lines, Fig.~\ref{fig_digitalgap_continents} also includes metrics regarding the model's (linear regression) goodness of fit, given by $R^2$ (where $R^2=1$ or 100\% means a perfect fit) and the RMSE in milliseconds. These two metrics indicate how accurate the regressed line is to the observed measurement points. As shown, the North America regressed line shows $R^2=0.738$, that is, the model explains 73.8\% of the dataset variability, and 12~ms of RMSE, which implies that the data points are very close to the regressed line (12~ms away of the linear fit on average), showing small variability (this is also observed in Fig.~\ref{fig:datapoints}). In these cases, the linear regression models provide nearly optimal parameters, with an intercept very close to 0~ms and a slope ver close to $0.01~ms/Km$. On the contrary, the same methodology applied to connected cities in Africa only explains 44.5\% of the variability and RMSE of 45~ms, since the measurement points are highly variable and show disparate distances to the regressed line whose parameters are also far from the optimal values. This further shows the difference in delay observed across regions and continents that tend to correlate with the wealth of countries in those regions.

\subsection{Latency measurements over the years: is delay reducing?}
\label{sec:year}

\begin{figure}[!t]
\centering
\includegraphics[width=0.95\columnwidth]{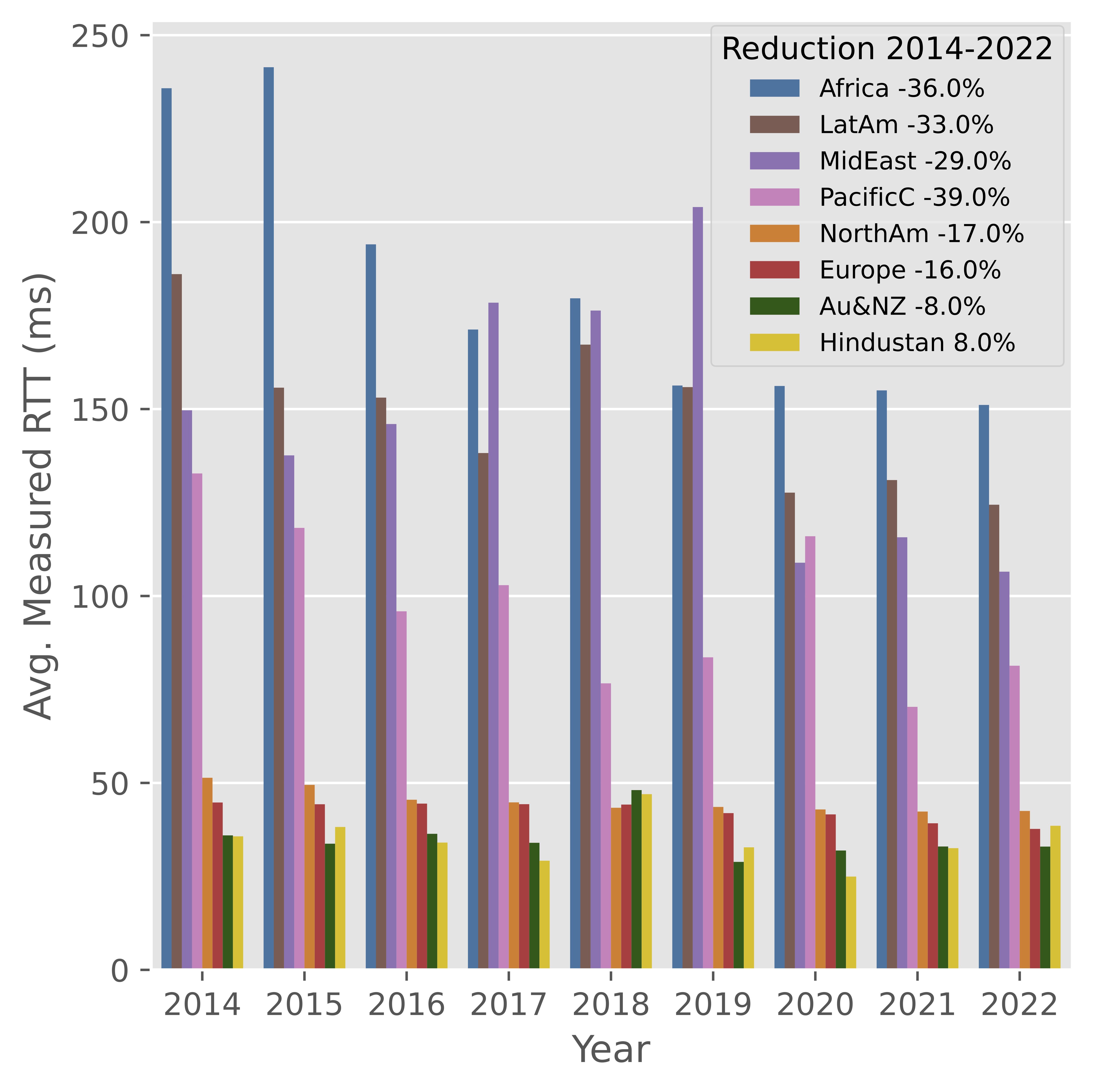}
\caption{Average RTT evolution over the years.}
\label{fig_evolution_over_years}
\end{figure}

Fig.~\ref{fig_evolution_over_years} shows average RTT values over the past eight years (2014-2022) for the eight sub-datasets. In most cases, network connectivity has experienced important latency improvements over the years, in spite of the increasing number of connected devices (especially things and mobile terminals) and applications emerged since 2014 (Social Networks, Video streaming platforms, etc). This evolution suggests that, although network traffic grows at a 30-35\% Compound Annual Growth Rate (CAGR), infrastructure investments made by telecom operators and countries keep reducing latency over the years. For instance, PacificC measurements show the largest latency reduction per year (39\% latency reduction in 8 years), from 135~ms average down to 80~ms. Europe and North America show a pace of 16\% and 17\% reduction respectively in 8 years, which translate into a rate of improvement of approximately 2.3\% latency reduction per year. 

\subsection{On delay variability: can latency bounds be guaranteed?}

\begin{figure}[htbp]

\centering
\subfigure[North America]{
\includegraphics[width=0.45\textwidth]{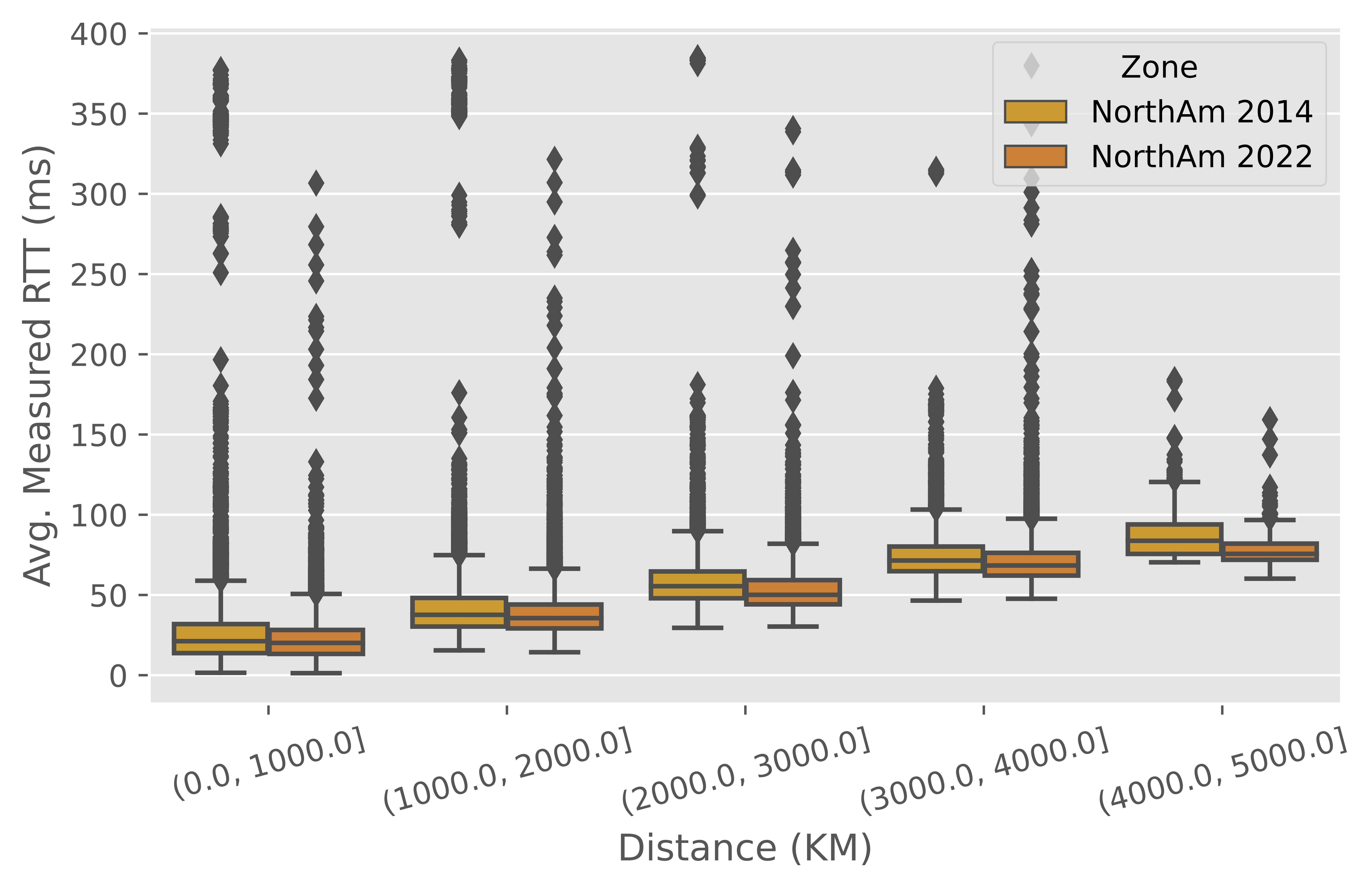}
\label{fig_latency_variability_a}
}
\subfigure[Latin America]{
\includegraphics[width=0.45\textwidth]{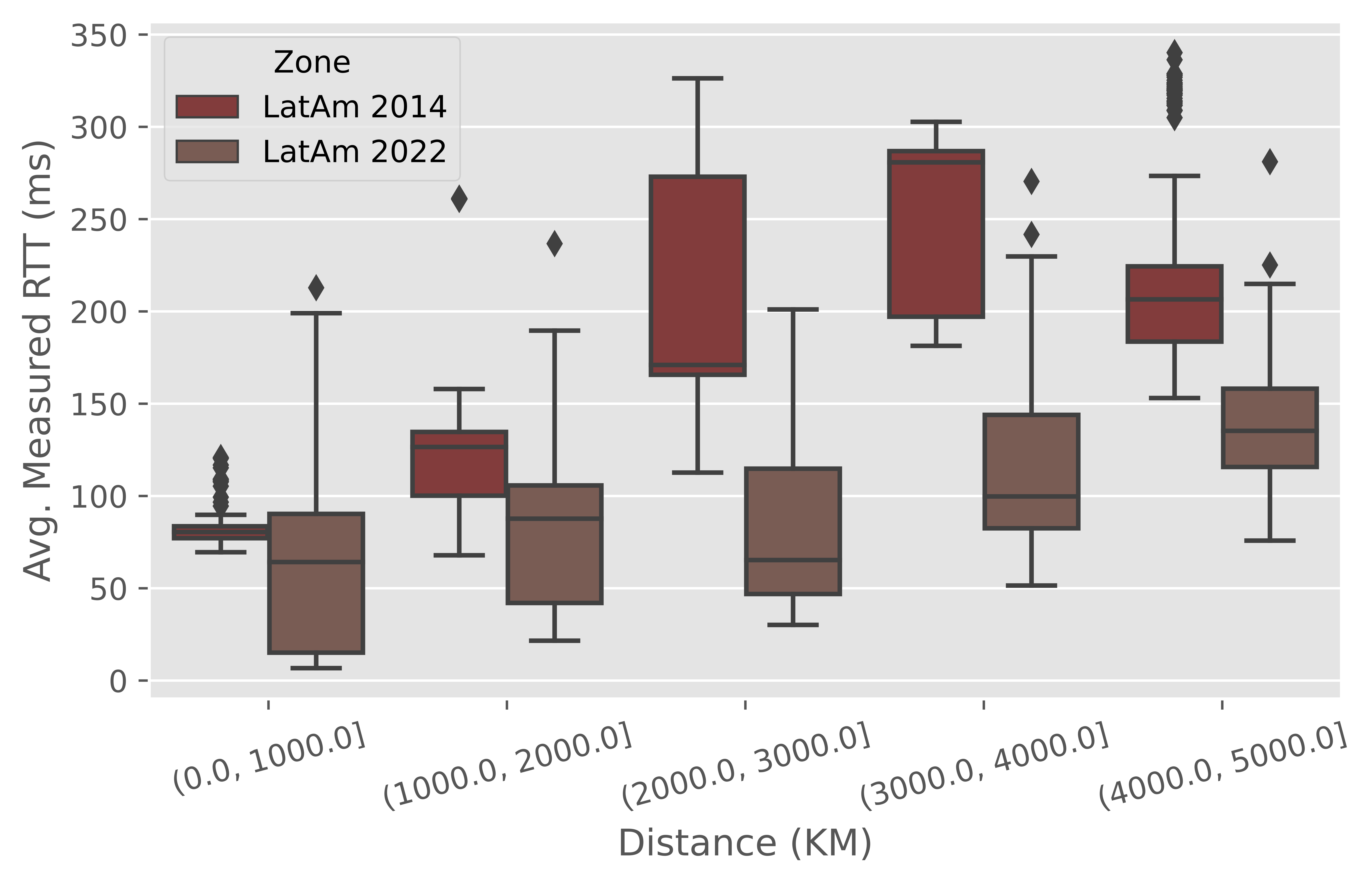}
\label{fig_latency_variability_b}
}
\caption{Latency and its variability over the years for NorthAm and LatAm.}
\end{figure}

Figs.~\ref{fig_latency_variability_a} and~\ref{fig_latency_variability_b} show boxplots of average RTT delays as obtained from measurements for both NorthAm and LatAm datasets respectively, between in 2014 and 2022 for different ranges of distances. Boxplots include the 25-th percentile (first quartile or Q1), 50-th percentile (i.e., median or second quartile, Q2), and 75-th percentile (third quartile, Q3). The whisker and dashed lines represent a distance of 1.5 times the interquartile range (IQR, Q3 $-$ Q1) above and below the third and first quartiles, respectively. All other points in the dataset outside the whiskers are considered and plotted as outliers (circles).

It is worth noticing that the vast majority of RTT values (those inside the whikers) are below 100 ms for NorthAm, while for LatAm the large majority of RTT values go up to 200~ms. The outliers in the NorthAm case correspond to cities not far away that experience high delay due to geographical constraints like mountains, lakes and deserts. The boxplots reveal that RTT delays have experienced important reductions not only in their median, but also the inter-quartile range between 2014 and 2022 and outliers, showing important latency variability reduction over the years, especially in North America.

\section{Applicability and use cases of RTT regression models}
\label{sec:applicability}

This section shows how the previous RTT regression models can be used to provide quick estimates of the RTT in two representative use cases. First, let us consider a company conducting its business in North America with three interconnected sites in the cities of Chicago, New York and Los Angeles. This forms a triangle with distances: 1,100~Km (Ch-NY), 2,800~Km (Ch-LA) and 3,900~Km (LA-NY). Following the NorthAm regression model for 2022 (i.e., RTT = 14.75 + 0.015 x Distance millisecs with RMSE = 12~ms), then we can safely estimate the average RTT values between the three cities as 31.25~ms (Ch-NY), 56.75~ms (Ch-LA) and 73.25~ms (LA-NY). This same distance example scenario for three separate cities in Africa (i.e., RTT = 61.85 + 0.022 x Distance millisecs with RMSE = 56~ms) would produce the following average RTT delays: 86.05~ms, 123.45~ms and 147.65~ms respectively, which is more than double.

In a second example, we have three friends enjoying an online videogame in London, Berlin and Madrid, but the server where the game is processed is located in Paris. Thus, the distances of interest are 340~Km for Lo-Pa, 880~Km for Be-Pa and 1,050~Km for Ma-Pa. Following the regression model for Europe (i.e, RTT = 10.89 + 0.02 x Distance with RMSE = 12~ms), we obtain 17.69~ms, 28.49~ms and 36.19~ms respectively. Same distances in other countries would probably make the online game not running smoothly. It is widely accepted that online gaming ping delays below 60~ms are considered sufficient, while values below 20~ms are considered ideal; values above 100~ms are often considered unacceptable for most gamers. Finally, it is worth mentioning that the access network (from the end-user's household to the aggregation and metro network, not included in the model) may comprise another 1-5~ms extra latency.

\section{Summary and discussion}
\label{sec:discussion}

This article evaluates latency experienced across monitoring stations located in major cities in the Internet worldwide. The measurement-based RTT analysis reveals multiple interesting observations:
 
\begin{itemize}
    \item There is a large technological gap (at least in terms of connectivity latency) between different countries and regions across the globe.
    \item Not only average latency, but also latency variability is also smaller in developed countries than emerging ones, which is an important key performance indicator for real-time applications.
    \item Some countries are indeed very close to the theoretical minimum, accounted as propagation delay between cities on straight fiber connections.
    \item Latency keeps reducing over the years in spite of annual traffic growths, showing that governments and telcos keep upgrading their networks steadily.
    \item Linear ML models obtained from real measurements provide both simple and powerful tools for network modelers and practitioners.
\end{itemize}

In general, it is important that countries continue investing on their networking infrastructure since this is a key enabler for the economy of the future, especially given the fact that traffic steadily grows 30-35\% per year. Forthcoming network-resource demanding services (i.e. connected cars, connected industries 4.0, Internet of things, 5G+/6G, metaverse, etc) will stress the network even further than today's popular services.

Emerging research activities on hollow-core fibers, where bits propagate at the speed of light (instead of two-thirds of it like in silica fibers), could also reduce network delay in the forth-coming years. Finally, the use of Low-Earth Orbit (LEO) satellites to interconnect those areas where fiber deployments are not possible may also allow important latency reductions. In addition, satellites benefit from smaller propagation delay than silica fibers (speed of light vs two-thirds of it).

\section*{Acknowledgment}

This work has been supported by the Spanish project ENTRUDIT (grant no. TED2021-130118B-I00) and the EU H2020 B5G-OPEN project (grant no. 101016663).

\bibliographystyle{IEEEtran}
\bibliography{Ping}

%








\begin{IEEEbiographynophoto}{Gonzalo Mart\'{i}nez} received his degree in Computer Engineering from Univ. Deusto (Bilbao, Spain), and the M.Sc. in Cybersecurity from Univ. Carlos III of Madrid (Spain) in 2019 and 2021 respectively. At present, he is currently conducting research in AI/ML techniques and data analytics and their applications in networking scenarios.
\end{IEEEbiographynophoto}

\begin{IEEEbiographynophoto}{Jos\'{e} Alberto Hern\'{a}ndez} completed his Degree Telecommunications Engineering at Univ. Carlos III de Madrid (Spain) in 2002, and Ph.D. Computer Science at Loughborough University (UK) in 2005. At present, he is with Univ. Carlos III de Madrid conducting teaching and research activities in Computer Networks, Optical WDM and AI/ML in networking.
\end{IEEEbiographynophoto}

\begin{IEEEbiographynophoto}{Pedro Reviriego} received the M.Sc. and Ph.D. degrees in telecommunications engineering from the Universidad Polit\'ecnica de Madrid, Spain, in 1994 and 1997, respectively. He is currently with Universidad Polit\'ecnica de Madrid working on probabilistic data structures, high-speed packet processing, and machine learning.
\end{IEEEbiographynophoto}

\begin{IEEEbiographynophoto}{Paul Reinheimer} completed the Bachelor of Computer Science, University of Windsor (Canada). In 2009, he co-founded WonderProxy to help his QA and development friends test their websites. WonderProxy has since grown from a weekend hack project to an actual business with servers and customers around the world.
\end{IEEEbiographynophoto}

\end{document}